\documentclass{iaus}

\usepackage{graphicx}

\title[Cepheids in IZw18]{Very metal poor Classical Cepheids: variables in IZw18.}

\author[short author list]   %% give here short author list %%
{G. Fiorentino $^1$, M. Marconi$^2$, \break G. Clementini
$^1$, I, Musella$^2$, A. Aloisi$^3$, F. Annibali$^3$,\\
 R. A. Contreras $^1$, \and M. Tosi $^1$}

\affiliation{$^1$INAF-OABo, via Ranzani 1, 40137, Bologna, Italia
\break email: giuliana.fiorentino@oabo.inaf.it\\[\affilskip]
$^2$ INAF-OAC, via Moiariello 16, 80131, Napoli, Italia \break\\[\affilskip]
$^3$ Space Telescope Science Institute, 3700 San Martin Drive,
Baltimore, MD 21218}

\pubyear{2004}

\volume{xxx}  %% insert here IAU Symposium No.

\pagerange{119--126}

\date{?? and in revised form ??}

\jname{Proceedings Title IAU Symposium}

\editors{A.C. Editor, B.D. Editor \& C.E. Editor, eds.}

\begin{document}

\maketitle

\begin{abstract}
In the framework of an ongoing ACS@HST project (HST program \#10586,
PI: Aloisi) we have obtained deep multi-color imaging of the very
metal-poor Blue Compact Dwarf galaxy IZw18. The data were acquired in
time-series fashion to allow the identification of Classical Cepheids
(CCs). The main aim of this project is to constrain both the distance
and the Star Formation History of the galaxy. However, as a byproduct
these data also provide new insights into the properties of CCs at
very low metallicities. We have identified 24 candidate CCs in
IZw18. New theoretical pulsation models of CCs specifically for the
low metallicity of this primordial galaxy (Z=0.0004, Y=0.24) have been
computed to interpret our results.

%The new models allow us to constrain the
%metallicity dependence of the Cepheid Period-Luminosity relation
%in the low metallicity regime (Z from 0.0004 to 0.008).



\end{abstract}

\firstsection % if your document starts with a section,


\section{Introduction}
\noindent The Blue Compact Dwarf galaxy IZw18 is one of the most intriguing
objects in the Local Universe. It has the lowest nebular metallicity
of all known star forming galaxies Z=1/30-1/50 Z$_\odot$ (depending on
the metallicity scale), and it has long been regarded as a possible
example of a galaxy undergoing its first burst of star formation,
hence a local analog of primordial galaxies in the distant
Universe. Previous HST observations with WFPC2 and NICMOS (Aloisi et
al., 1999, Ostlin et al., 2000) detected AGB stars tracing an
underlying population about 0.5 Gyr old. However, Izotov \& Thuan,
(2004, IT04) failed to reveal the galaxy RGB tip on their ACS@HST
data, thus suggesting that IZw18 does not host an old population.
Since current estimates for the distance of IZw18 are very uncertain,
it may be possible that IZw18 RGB tip was not detected because the
galaxy is farther away than previously thought.  We have used our
ACS@HST deep time-series data of IZw18, to measure the galaxy distance
with the CCs.

\section{Data reductions \& Theoretical models}\label{sec:obs}
\noindent The proprietary ACS@HST data were combined with IT04 images,
reaching a total time baseline of about 2.5 years. Photometric
reductions were performed using the ALLFRAME package (Stetson 1987,
1992).  Variable stars were detected using both the Optimal Image
Subtraction Technique (ISIS 2.2, Alard 2000) and the Welch \& Stetson
(1993) variability index. We have identified 24 variable star
candidates (triangles in panel a), Fig. 1). Analysis of the light
curves is in progress.  The $V,I$ light curves of a CC with period of
8.8 d (star \#4306) are shown in Fig. 1 (panel b).

\begin{figure}
\includegraphics[height=5.5cm,width=6.5cm]{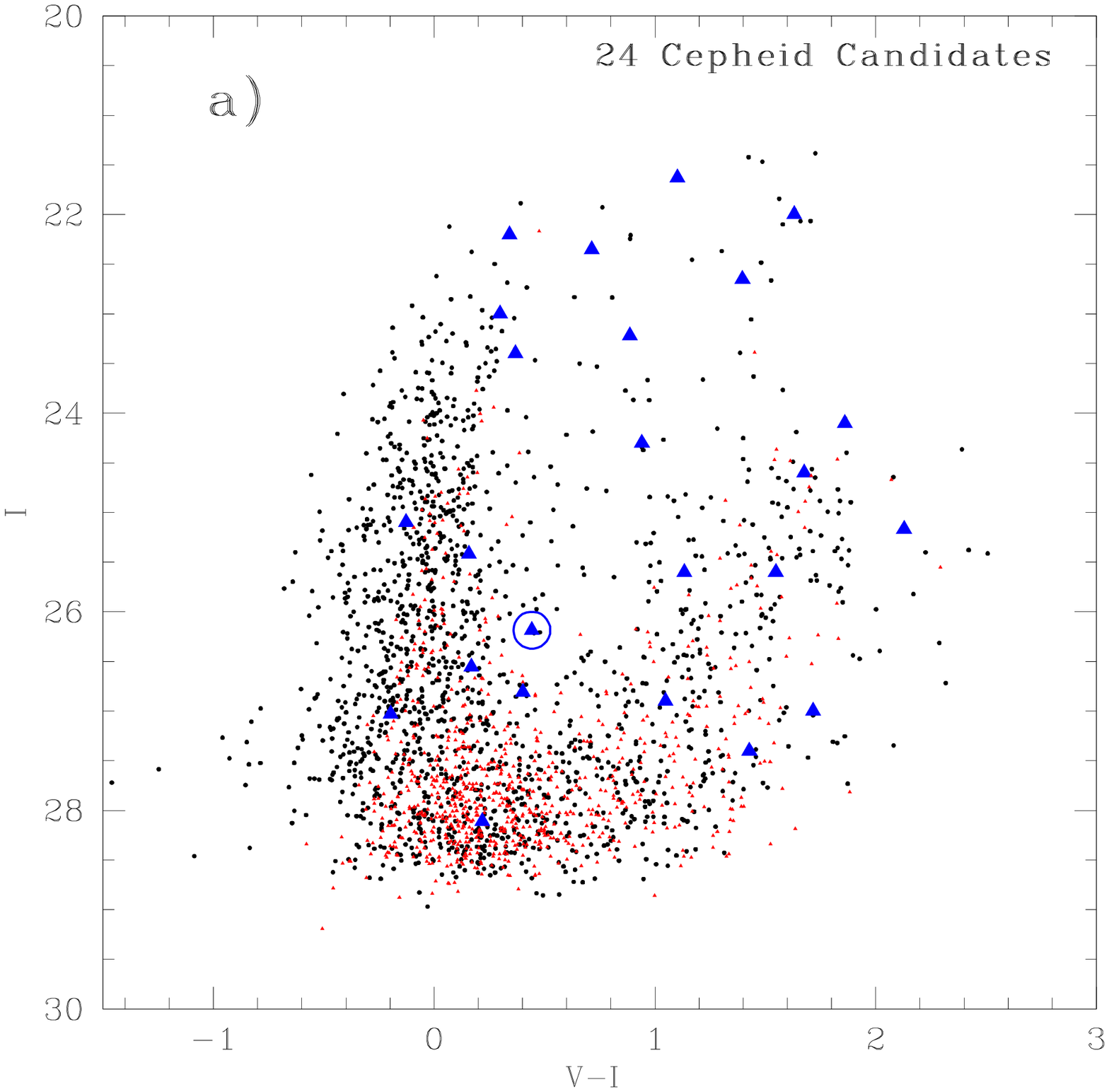}
\includegraphics[height=5.5cm,width=6.5cm]{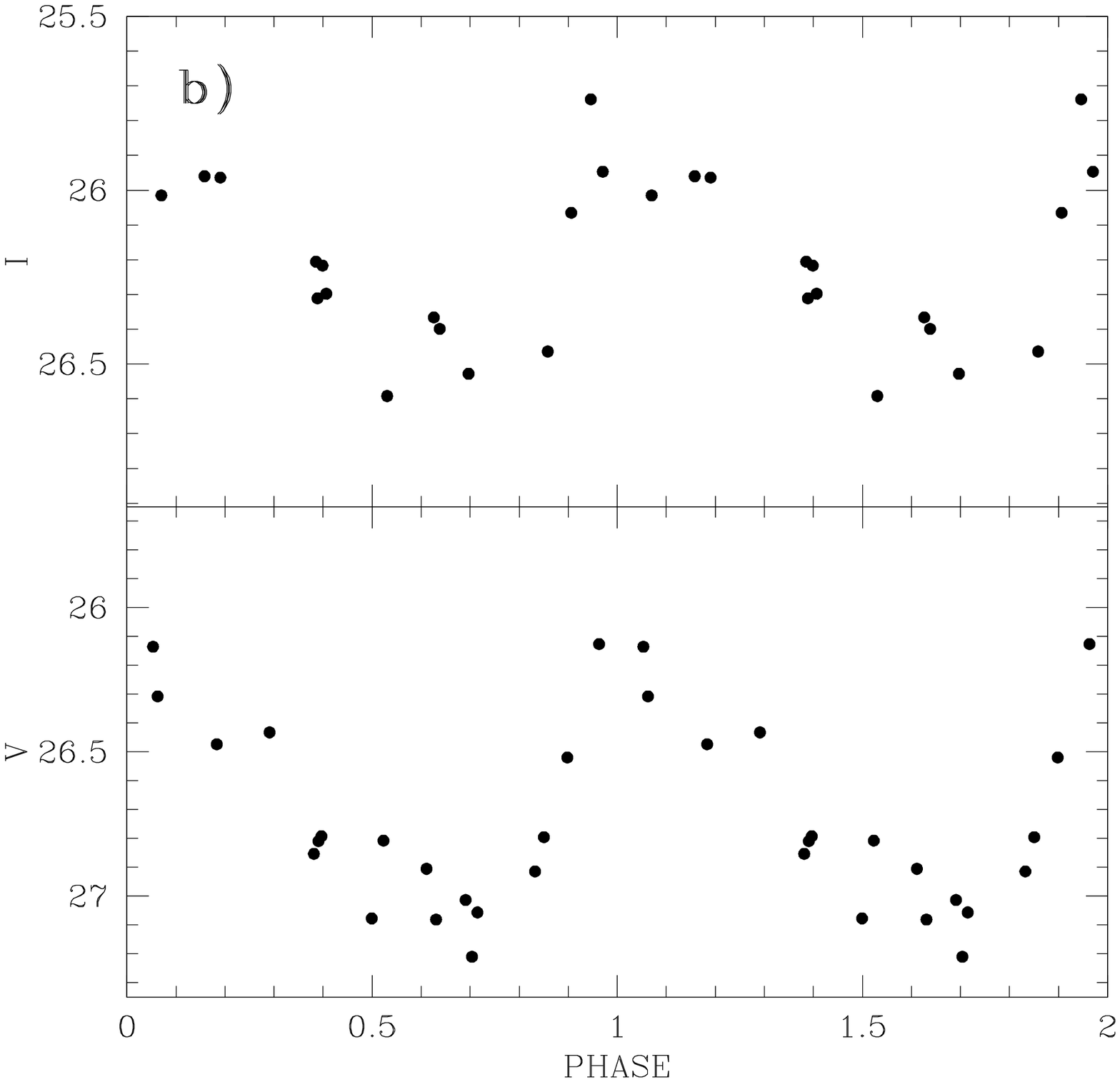}
  \caption{ Panel a): CMD of IZw18 obtained combining F555W, F606W
and F814W ACS data from IT04 and from Aloisi et al. HST program
\#10586. Candidate variables are marked by filled triangles. Panel
b): $V,I$ light curves for a P=8.8 d CC (see open circles in
panel a).
%Star ID and period are given in the top-label.
    }\label{cmd}
\end{figure}

\noindent New theoretical nonlinear pulsation models of CCs at the metallicity
of IZw18 (Z=0.0004, Y=0.24) have been computed, (see Marconi et al. 2005 for
details on the pulsation code) to constrain the metallicity dependence
of the Cepheid Period-Luminosity (PL) relation in the low metallicity
regime. The effect of different assumptions in the mixing length
parameter (l/HP=1.5-2) on the classical pulsation relations (PLC,
Wesenheit and PL) and on the theoretical light curves has been
investigated.  At this low metallicity, the effect of increasing the
convection efficiency is negligible on the light curve amplitude, but
causes a decrease in effective temperature width of the Instability
Strip. The metallicity effect on the PL relations is negligible in
this metal-poor range.

\section{Conclusions}\label{sec:obs}

\noindent The distance modulus of IZw18 was determined by applying both the
theoretical and the empirical approach to our most reliable CCs
(namely stars: \#8431 and 4306): 1) Our theoretical Wesenheit relation
(Fiorentino et al. 2007, in preparation) gives : $\mu_0$ =31.28 $\pm$
0.26 mag; 2) Ogle's empirical PL-relation for the LMC (Udalski et
al. 1999): gives  $\mu_0$ = 30.71 $\pm$ 0.29 mag, for an assumed distance
modulus for the LMC of 18.5 mag. If we take into account the
metallicity correction, we find: 1) a decrease of the distance modulus
to: $\mu_0$=30.40 $\pm$ 0.29 mag if the empirical correction by
Kennicut et al. (1998) is adopted; 2) an increase of the distance
modulus to $\mu_0$ = 31.01 $\pm$ 0.29 mag, if we extrapolate to the
metallicity of IZw18 the theoretical correction by Fiorentino et
al. (2002).


\begin{thebibliography}{}

\bibitem[Aloisi et al. (1999)]{a99}{Aloisi, A., Tosi, M. \& Greggio, L., 1999, AJ, 118, 302A}
\bibitem[Fiorentino et al.(2002)]{f02}{Fiorentino, G., Caputo, F., Marconi, M., Musella, I., 2002, ApJ, 576,
402}
\bibitem[Kennicutt, et al. (1998)]{k98}{Kennicutt, R. C., Jr., et al. 1998, ApJ, 498, 181}
\bibitem[Izotov \& Thuan (2004)]{it04}{Izotov, Y. I. \& Thuan, T. X., 2004, ApJ, 616, 768I}
\bibitem[Marconi et al.(2005)]{m05}{Marconi, M., Musella, I., Fiorentino, G., 2005, ApJ, 632, 590M}
\bibitem[Ostlin, G. (2000)]{o00}{Ostlin, G., 2000, ApJ, 535L, 99O}
\bibitem[Welch \& Stetson (1993)]{ws93}{Welch, D. L. \& Stetson, P. B., 1993, AJ, 105, 1813W}
\bibitem[Stetson (1987)]{s87}{Stetson, Peter B., 1987, PASP, 99, 191S}
\bibitem[Stetson (1992)]{s92}{Stetson, Peter B., 1992, ASPC, 25, 297S}
\bibitem[Udalski et al. (1999)]{u99}{Udalski, A., et al. 1999, Acta Astron., 49,
223}

\end{thebibliography}
\end{document}